\documentclass[10pt,showpacs,amsmath,amssymb,floatfix,superscriptaddress,longbibliography,twocolumn,eqsecnum]{revtex4-2}
\usepackage{amsmath}
\usepackage{physics}
\usepackage{graphicx}
\usepackage[usenames,dvipsnames,svgnames,table]{xcolor}
\usepackage{tcolorbox}
\usepackage{booktabs}
\usepackage{makecell} 
\usepackage{tabularx}
\usepackage{chngcntr}
\usepackage[utf8]{inputenc}
\counterwithout{equation}{section}
\counterwithout{figure}{section}
\usepackage[unicode=true,
            pdfusetitle,
            bookmarks=true,
            bookmarksnumbered=false,
            bookmarksopen=false,
            breaklinks=true,
            pdfborder={0 0 0},
            backref=false,
            colorlinks=true,
            hypertexnames=false]{hyperref}
\hypersetup{linkcolor=NavyBlue,urlcolor=NavyBlue,citecolor=NavyBlue}

\begin{document}

\title{Correlation-Assisted Odd-Parity Encoded Gates in Coupled Fluxonium Qubits under Non-Markovian TLS Noise}

\author{Chenghong Ji}
\affiliation{School of Physics, Hangzhou Dianzi University, Hangzhou 310018, China}
\author{Chaoying Zhao}
\email{zchy49@163.com}
\affiliation{School of Physics, Hangzhou Dianzi University, Hangzhou 310018, China}
\affiliation{State Key Laboratory of Quantum Optics Technologies and Devices, Shanxi University, Taiyuan, 030006,China}
\affiliation{Zhejiang Key Laboratory of Quantum State Control and Optical Field Manipulation, Hangzhou Dianzi University, Hangzhou, 310018, China}
\date{\today}

\begin{abstract}
Correlated longitudinal noise can be partially converted into common-mode fluctuations in an odd-parity two-qubit subspace. We analyze an encoded logical qubit formed by the states $\ket{0_L}=\ket{01}$ and $\ket{1_L}=\ket{10}$ in two coupled fluxonium qubits. Projecting the exchange-coupled two-qubit Hamiltonian onto this subspace yields an effective logical Hamiltonian in which the exchange interaction drives $X_L$ rotations and the qubit detuning drives $Z_L$ rotations. We model correlated two-level-system (TLS) noise by using longitudinal stochastic processes with finite memory time and evaluate encoded-gate performance through the average gate fidelity. Within the projected model, positive spatial noise correlation suppresses the differential fluctuation and thereby improves the fidelity of encoded logical gates. We further compare Gaussian Ornstein-Uhlenbeck, Markovian, and random-telegraph noise models and examine the role of logical dynamical decoupling. These results identify a noise-adapted control mechanism for odd-parity encoded operations in coupled fluxonium devices and motivate future multilevel simulations including leakage and pulse-level constraints.
\end{abstract}

\maketitle

\section{Introduction}
Recent progress in fluxonium circuits has made this modality a serious candidate for gate-based superconducting quantum information processing. Compared with weakly an-harmonic transmons, fluxonium combines large anharmonicity, low transition frequency, and long coherence, which are favorable for suppressing leakage and designing spectrally selective gates. Millisecond-scale coherence, scalable processor blueprints, and high-fidelity two-fluxonium gates based on tunable, inductive, direct, and resonator-mediated couplers have been demonstrated or proposed in recent experiments and designs \cite{Bao2022,Somoroff2023,Nguyen2022,Moskalenko2022,Zhang2024,Rosenfeld2024,Mencia2024,Lin2025}. These results suggest that fluxonium is not only a protected physical qubit, but also a useful platform for exploring hardware-adapted logical operations.

However, the same circuit features that protect fluxonium from some noise channels also introduce a complex materials environment. In particular, Josephson-junction chains and amorphous interfaces can host two-level-system defects, whose slow dynamics and discrete spectral structure may invalidate a purely Markovian description \cite{deGraaf2021,Abdurakhimov2022,Chen2024,Odeh2025,Zhuang2026}. Recent non-Markovian studies of superconducting qubits show that memory effects can be reconstructed, modeled with post-Markovian or temporally correlated noise descriptions, and connected to gate-level errors \cite{Zhang2022,White2022,Gulacsi2023,Zou2024}. For two coupled qubits, the crucial issue is not only the strength of the TLS noise, but also its spatial correlation. A common-mode longitudinal fluctuation produces nearly the same phase shift on the two physical qubits, whereas the odd-parity coherence between $\ket{01}$ and $\ket{10}$ depends mainly on their relative phase.

This observation motivates a shift from passive entanglement protection to active logical-gate construction. Decoherence-free and noise-adapted encodings have suggested that symmetry in the noise can be converted into a computational resource \cite{Zanardi1997,Lidar1998}. In fluxonium devices, the relevant symmetry can arise from correlated TLS-induced frequency fluctuations. Instead of encoding information in one physical qubit, we define $\ket{0_L}=\ket{01}$ and $\ket{1_L}=\ket{10}$ within the odd-parity subspace. The exchange coupling between the two fluxoniums then provides an effective $X_L$ operation, while qubit detuning provides a controllable $Z_L$ operation.

In our previous work, we have analyzed two capacitively coupled fluxonium qubits under correlated Ornstein-Uhlenbeck TLS noise and showed that odd-parity entanglement can be extended by correlation-assisted protection and TLS-optimized dynamical decoupling~\cite{ji2026}. The present work goes beyond state preservation and asks whether the same correlated-noise
structure can be converted into a gate-level advantage for encoded logical control. First, we formulate an odd-parity encoded logical qubit in two coupled fluxonium qubits subject to correlated non-Markovian TLS noise. Second, we derive the projected logical Hamiltonian and identify exchange- and detuning-based logical control axes. Third, we benchmark encoded logical gates by using average gate fidelity and leakage probability \cite{ABAD2022}, thereby connecting microscopic correlated TLS physics to programmable logical operations in fluxonium superconducting processors.

\section{Theoretical Model}

To identify a noise-adapted logical subspace in two coupled fluxonium qubits, we first introduce a fluxonium-motivated two-qubit model and then project it onto the odd-parity manifold spanned by $\ket{01}$ and $\ket{10}$. This projection separates longitudinal TLS-induced frequency fluctuations into a common-mode component and a differential component. The common-mode component contributes only an overall phase within the odd-parity subspace, whereas the differential component produces logical dephasing. This common-mode--differential-mode separation is the central mechanism underlying the correlation-assisted encoded gates analyzed below.

\begin{figure}[htbp]
	\centering
	\includegraphics[width=8.6cm]{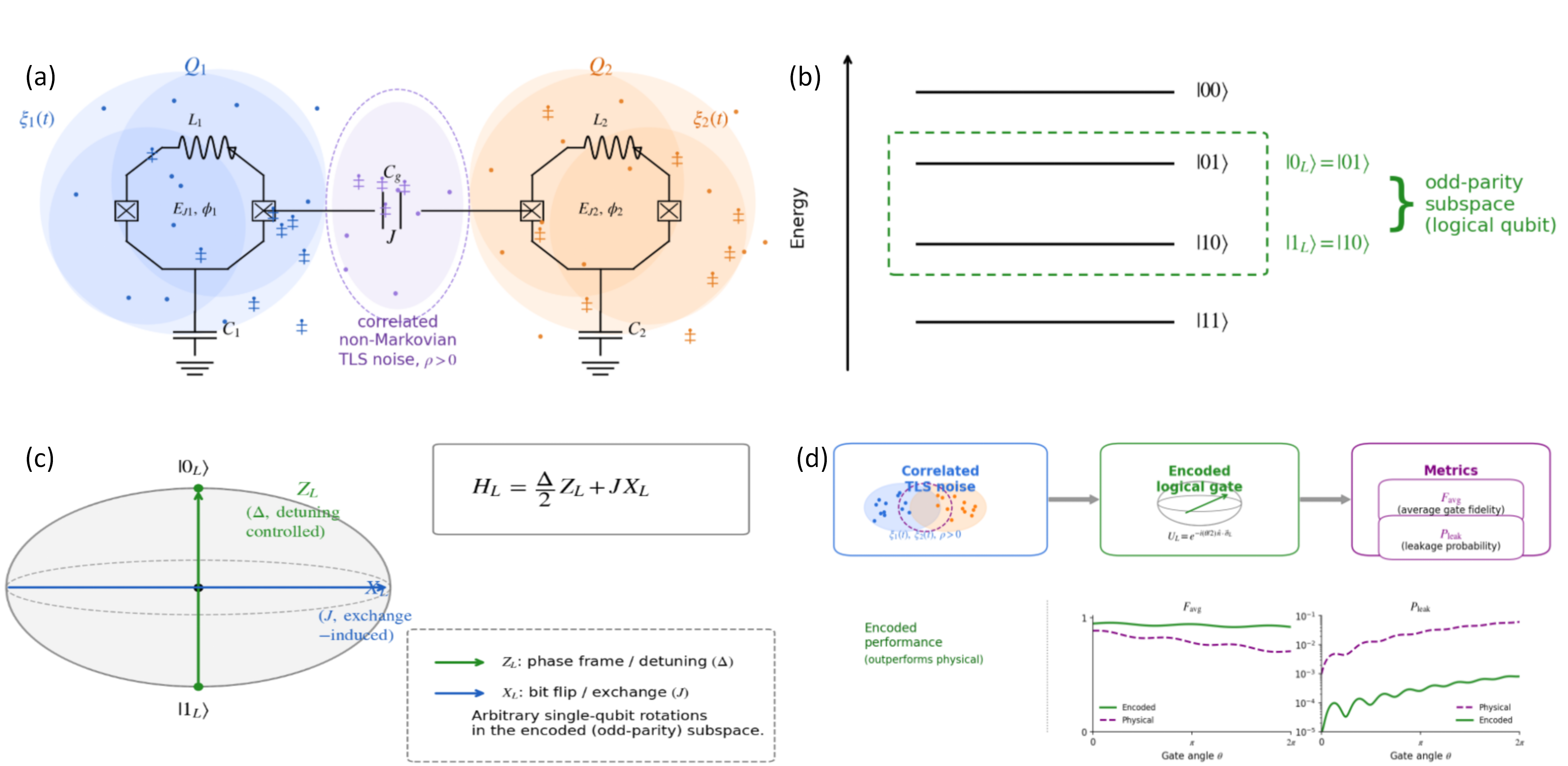}
	\caption{Physical picture of correlation-assisted odd-parity encoded control.
		(a) Two coupled fluxonium qubits are exposed to partially correlated longitudinal TLS-induced fluctuations. The overlap of the local defect environments produces a common fluctuating component characterized by the spatial correlation coefficient $\rho$.
		(b) The logical qubit is encoded in the odd-parity subspace, $\ket{0_L}=\ket{01}$ and $\ket{1_L}=\ket{10}$.
		(c) Projection onto this subspace yields an effective logical Hamiltonian in which exchange coupling generates $X_L$ rotations and qubit detuning generates $Z_L$ rotations.
		(d) In the projected logical model, noisy encoded operations are compared with ideal target gates by using the average gate fidelity. Leakage is absent by construction in the projected model.
	}
	\label{fig:1}
\end{figure}

Fig.~\ref{fig:1} summarizes the physical picture considered in this work. We consider two capacitively coupled fluxonium qubits, denoted by $Q_1$ and $Q_2$, which are subject to partially correlated longitudinal fluctuations generated by TLS defects. Each fluxonium consists of a Josephson element, a super-inductance, and a shunting capacitance. The two qubits are coupled through a small capacitance, giving rise to an effective exchange interaction $J$ after projection onto the qubit subspace. The overlap of the local TLS environments produces a common fluctuating component, characterized by the spatial correlation coefficient $\rho$. The goal of the model is to capture how this correlated longitudinal noise acts after projection into the odd-parity logical subspace.

At the circuit level, the Hamiltonian of two capacitively coupled fluxonium qubits can be written as \cite{manucharyan2009}
\begin{equation}
	\begin{aligned}
		H_{\mathrm{circuit}} &=
		\sum_{i=1}^{2}
		\left(
		4E_{C_i}\hat{n}_i^2
		+\frac{1}{2}E_{L_i}
		\left(\hat{\varphi}_i-\varphi_{\mathrm{ext},i}\right)^2
		-E_{J_i}\cos\hat{\varphi}_i
		\right)  \\
		&\quad
		+8E_{C_{12}}\hat{n}_1\hat{n}_2
		+H_{\mathrm{TLS}}
		+H_{\mathrm{q-TLS}}
	\end{aligned}
	\label{eq:circuit_hamiltonian}
\end{equation}

Here, $\hat{\phi}_i$ and $\hat{n}_i$ are the superconducting phase and Cooper-pair number operators of the $i$-th fluxonium, satisfying $[\hat{\phi}_i,\hat{n}_j]=i\delta_{ij}$. The parameters $E_{C_i}$, $E_{J_i}$, and $E_{L_i}$ denote the charging, Josephson, and inductive energies, respectively. The capacitive term $8E_{C_{12}}\hat{n}_1\hat{n}_2$ originates from the coupling capacitance and gives rise to an effective exchange interaction after projection onto the lowest two levels of each fluxonium.

To connect the effective two-qubit description with the underlying circuit, each isolated fluxonium Hamiltonian can be diagonalized in a truncated oscillator basis
\begin{equation}
	H_i^{fl}
	=
	4E_{C i}\hat n_i^2
	+
	\frac{1}{2}E_{L i}
	\left(
	\hat\varphi_i-\varphi_{{\rm ext},i}
	\right)^2
	-
	E_{J i}\cos\hat\varphi_i .
	\label{eq:single_fluxonium}
\end{equation}
After diagonalization, it becomes
\begin{equation}
	H_i^{fl}
	=
	\sum_{m=0}^{N_{\rm lev}-1}
	\epsilon_m^{(i)}
	|m_i\rangle\langle m_i| ,
	\label{eq:single_fluxonium_diag}
\end{equation}
where $|m_i\rangle$ is the $m$-th eigen-state of the $i$-th fluxonium. We define the qubit frequency and anharmonicity are 
\begin{equation}
	\omega_{01}^{(i)}
	=
	\frac{\epsilon_1^{(i)}-\epsilon_0^{(i)}}{\hbar},
	\qquad
	\alpha_i
	=
	\omega_{12}^{(i)}-\omega_{01}^{(i)} ,
	\label{eq:freq_anharm}
\end{equation}
with
\begin{equation}
	\omega_{12}^{(i)}
	=
	\frac{\epsilon_2^{(i)}-\epsilon_1^{(i)}}{\hbar}.
\end{equation}

The projected two-level description is valid when the exchange coupling and detuning used for logical control are small compared with the fluxonium anharmonicity \cite{manucharyan2009fluxonium}
\begin{equation}
	|J|,|\Delta|\ll |\alpha_i| ,
\end{equation}
which justifies the projected two-level description for the encoded dynamics.

After diagonalization of the multilevel fluxonium Hamiltonian, we project the dynamics onto the two lowest eigen-states of each device. This projection is justified by the large fluxonium anharmonicity, $|\alpha_i|/2\pi\sim 5~{\rm GHz}$, compared with the exchange coupling and detuning used for logical control \cite{blais2021circuit}.

\begin{equation}
	H_0
	=
	\frac{\hbar\omega_1}{2}\sigma_z^{(1)}
	+
	\frac{\hbar\omega_2}{2}\sigma_z^{(2)}
	+
	\hbar J
	\left(
	\sigma_+^{(1)}\sigma_-^{(2)}
	+
	\sigma_-^{(1)}\sigma_+^{(2)}
	\right)
	\label{eq:two_qubit_hamiltonian}
\end{equation}
where $\omega_i$ is the transition frequency of $Q_i$, and $J$ describes the coherent exchange of a single excitation between the two fluxonium qubits.

The TLS environment is modeled as a correlated longitudinal frequency fluctuation acting on the two physical qubits \cite{ithier2005decoherence}:
\begin{equation}
	H_{\mathrm{noise}}(t)
	=
	\frac{\hbar}{2}
	\sum_{i=1}^{2}
	\xi_i(t)\sigma_z^{(i)} 
	\label{eq:noise_hamiltonian}
\end{equation}
The stochastic variables $\xi_i(t)$ corresponds to zero-mean Gaussian Ornstein-Uhlenbeck processes with correlation function \cite{wang1945theory}
\begin{equation}
	\left\langle
	\xi_i(t)\xi_j(t')
	\right\rangle
	=
	\sigma_i\sigma_j\rho_{ij}
	e^{
	-\frac{|t-t'|}{\tau_c}}
	\label{eq:ou_correlation}
\end{equation}
and corresponding Lorentzian spectrum
\begin{equation}
	S_{ij}(\omega)
	=
	\frac{2\sigma_i\sigma_j\rho_{ij}\tau_c}
	{1+\omega^2\tau_c^2}
	\label{eq:lorentzian_spectrum}
\end{equation}
Here, $\tau_c$ is the TLS memory time, $\sigma_i$ is the noise amplitude, and $\rho_{ij}$ quantifies the spatial correlation of the TLS noise. In particular, $\rho_{12}=\rho>0$ corresponds to the overlapping TLS region in Fig.~1(a).

We use the odd-parity subspace as a logical qubit. As shown in Fig.1(b), we define
\begin{equation}
	|0_L\rangle = |01\rangle
	\qquad
	|1_L\rangle = |10\rangle 
	\label{eq:logical_basis}
\end{equation}
The projector onto the logical subspace is
\begin{equation}
	P_L
	=
	|01\rangle\langle 01|
	+
	|10\rangle\langle 10| 
	\label{eq:logical_projector}
\end{equation}
Within this subspace, the logical Pauli operators are
\begin{equation}
	X_L
	=
	|01\rangle\langle 10|
	+
	|10\rangle\langle 01|
	\label{eq:logical_x}
\end{equation}
\begin{equation}
	Z_L
	=
	|01\rangle\langle 01|
	-
	|10\rangle\langle 10| 
	\label{eq:logical_z}
\end{equation}

Projecting Eq.~\eqref{eq:two_qubit_hamiltonian} into the logical subspace
\begin{equation}
	H_L
	=
	P_L H_0 P_L
	=
	\hbar
	\left(
	\frac{\Delta}{2}Z_L
	+
	JX_L
	\right)
	\label{eq:logical_hamiltonian}
\end{equation}
where
\begin{equation}
	\Delta=\omega_1-\omega_2 
	\label{eq:detuning}
\end{equation}
Eq.~\eqref{eq:logical_hamiltonian} provides the control mechanism shown in Fig.~1(c): the exchange coupling $J$ generates logical $X_L$ rotations, the detuning $\Delta$ generates logical $Z_L$ rotations. Therefore, a general single-logical-qubit rotation can be implemented as
\begin{equation}
\begin{split}
	U_L(T)
	&=e^
	{-\frac{i}{\hbar}
	H_L T} =
	e^{
	-iT
	\left(
	\frac{\Delta}{2}Z_L
	+
	JX_L
	\right)}
	\label{eq:logical_gate}
\end{split}
\end{equation}
For example, when $\Delta=0$, the exchange term realizes a $X_L$ rotation. When $J$ is effectively suppressed or much smaller than $|\Delta|$, the detuning term realizes a $Z_L$ rotation.

The projected noise Hamiltonian is
\begin{equation}
\begin{split}
	&H_{\mathrm{noise}}^{(L)}(t)
	= P_L H_{\mathrm{noise}}(t)P_L \\
	&= \frac{\hbar}{2} \left(\xi_1(t)-\xi_2(t) \right) Z_L
	+ \frac{\hbar}{2} \left(\xi_1(t)+\xi_2(t) \right) \mathbb{I}_L 
\end{split}
\label{eq:logical_noise}
\end{equation}

The second term contributes a common phase in the logical subspace, the first term causes logical dephasing.

\begin{equation}
	\delta\xi(t)=\xi_1(t)-\xi_2(t)
	\label{eq:differential_noise}
\end{equation}
Its variance is reduced by positive noise correlation:
\begin{equation}
	\left\langle
	\delta\xi(t)\delta\xi(t')
	\right\rangle
	=
	\left(
	\sigma_1^2+\sigma_2^2-2\rho\sigma_1\sigma_2
	\right)
	e^{
	-\frac{|t-t'|}{\tau_c}}
	\label{eq:differential_correlation}
\end{equation}
For the symmetric case, $\sigma_1=\sigma_2=\sigma$,
\begin{equation}
	\left\langle
	\delta\xi(t)\delta\xi(t')
	\right\rangle
	=
	2\sigma^2(1-\rho)
	e^{
	-\frac{|t-t'|}{\tau_c}}
	\label{eq:symmetric_noise}
\end{equation}

\begin{table}[t]
	\centering
	\caption{
		Representative parameters used for the projected encoded-gate simulations. The values correspond to a heavy-fluxonium regime with low qubit frequency, large anharmonicity, and MHz-scale exchange and detuning. The logical detuning is defined as $\Delta=\omega_1-\omega_2$. The listed $Z_L(\pi)$ gate time assumes that the exchange contribution is suppressed, echo-compensated, or treated in an appropriate rotating frame during the detuning-based operation.
	}
	\label{tab:parameters}
	\begin{tabular}{|l|c|c|}
		\hline
		\textbf{Parameter} & \textbf{Symbol} & \textbf{Value} \\
		\hline
		Charging energy       & $E_C/h$            & $1.0$    $GHz$ \\
		Josephson energy      & $E_J/h$            & $7.0$    $GHz$ \\
		Inductive energy      & $E_L/h$            & $0.50$   $GHz$ \\
		External flux bias    & $\varphi_{\rm ext}$ & $\simeq \pi$ \\
		Qubit frequency, $Q1$   & $\omega_{01}^{(1)}/2\pi$ & $50$  $MHz$ \\
		Qubit frequency, $Q2$   & $\omega_{01}^{(2)}/2\pi$ & $55$  $MHz$ \\
		Detuning              & $|\Delta|/2\pi$    & $5$      $MHz$ \\
		Anharmonicity         & $|\alpha|/2\pi$    & $\sim 5$ $GHz$ \\
		Exchange coupling     & $J/2\pi$           & $2.5$    $MHz$ \\
		$X_L(\pi)$ gate time  & $T_X^{(\pi)}$      & $100$    $ns$ \\
		$Z_L(\pi)$ gate time  & $T_Z^{(\pi)}$      & $100$    $ns$ \\
		Noise amplitude       & $\sigma/2\pi$      & $0.30$   $MHz$ \\
		TLS memory time       & $\tau_c$           & $1.0$    $\mu s$ \\
		Correlation coefficient & $\rho$           & $0-0.99$ \\
		\hline
	\end{tabular}
\end{table}

The logical control scales satisfy
\begin{equation}
	\frac{|J|}{|\alpha|}
	\sim
	5\times 10^{-4},
	\qquad
	\frac{|\Delta|}{|\alpha|}
	\sim
	10^{-3}
	\label{eq:small_parameter_ratios}
\end{equation}
These small ratios justify the use of the projected odd-parity Hamiltonian in Eq.~\eqref{eq:logical_hamiltonian} as an effective description of the logical dynamics. Within this projected model, leakage outside the logical subspace is absent by construction. The simulations quantify logical-channel errors caused by residual differential dephasing and coherent control imperfections. A complete leakage analysis requires a multi-level fluxonium simulation including higher excited states, pulse shapes, transverse noise components, and non-adiabatic effects.

Eqs.(\ref{eq:logical_noise})-(\ref{eq:symmetric_noise}) show that positive spatial correlation suppresses the differential longitudinal noise experienced by the encoded logical qubit. In the limiting $\rho\rightarrow 1$, the common-mode TLS fluctuation contributes a global phase, while logical dephasing from $\delta\xi(t)$ is strongly reduced. This correlation-assisted suppression provides a physical basis for construct encoded logical gates in the odd-parity subspace.

Table~\ref{tab:parameters} summarizes the representative device, control, and noise parameters used in the numerical analysis. These values are chosen to describe a heavy-fluxonium regime with low qubit transition frequencies, large anharmonicity, and $MHz$-scale logical control. We set the scale of the projected encoded-gate simulations rather than representing a full calibrated multilevel device model.

The high-level workflow in Fig.~\ref{fig:1}(d) indicates how this encoded logical model is bench-marked: the noisy encoded operation  compared with the ideal target gate, and the performance is quantified by average gate fidelity $F_{\rm avg}$ and leakage probability $P_{\rm leak}$.

\section{Encoded Logical Gate Construction and Benchmarking}

After establishing the correlation-assisted protection mechanism of the odd-parity logical subspace, we turn to encoded logical-gate construction. The projected Hamiltonian in Eq.~(\ref{eq:logical_hamiltonian}) provides two independent logical control axes: the exchange coupling $J$ generates $X_L$ rotations, the qubit detuning $\Delta$ generates $Z_L$ rotations. This section connects these control axes to elementary single-logical-qubit gates and defines the benchmarking protocol used in the numerical simulations.

Unlike state-preservation analysis, logical-gate characterization must compare a noisy implemented operation with a target unitary operation. Therefore, we track both the coherent rotation within the encoded subspace and the population that may leak outside the logical manifold in realistic fluxonium implementations.

\begin{figure}[htbp]
	\centering
	\includegraphics[width=8.6cm]{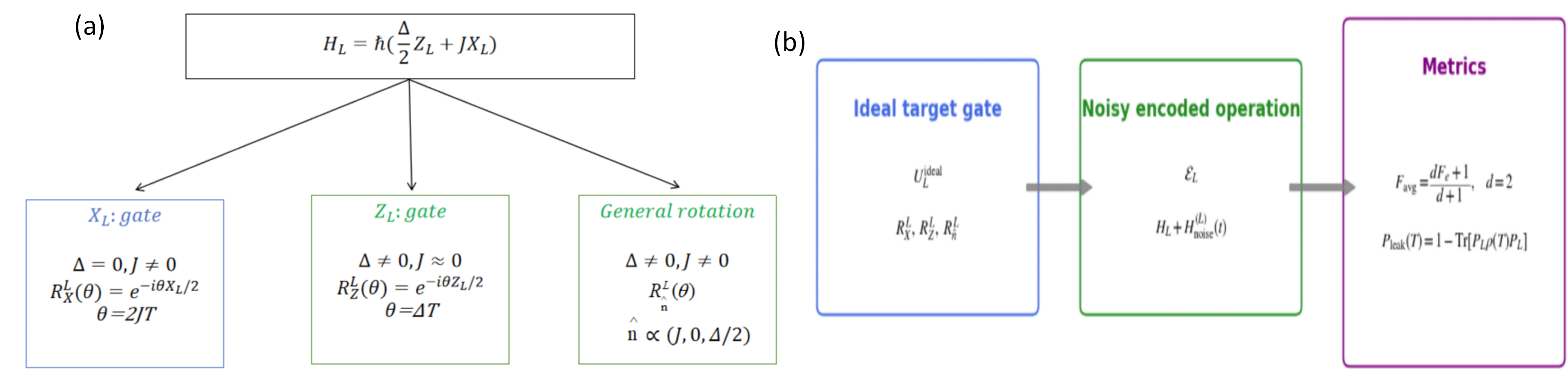}
	\caption{
		Construction and benchmarking of odd-parity encoded logical gates. (a)Logical rotations generated by the projected Hamiltonian $H_L=\hbar((\Delta/2)Z_L+JX_L)$. The exchange-only limiting implements a $X_L$ rotation, an exchange-compensated detuning operation implements a $Z_L$ rotation. When both $J$ and $\Delta$ are present during the same gate window, the logical qubit undergoes a tilted-axis rotation. (b) Benchmarking protocol in the projected logical model. For each stochastic realization, the noisy logical propagator compared with a specified target unitary, and the logical quantum channel is evaluated by using the average gate fidelity $F_{\rm avg}$. Leakage outside the odd-parity manifold is not included in the projected simulations.
		}
	\label{fig:2}
\end{figure}

As shown in Fig.~\ref{fig:2}(a), we start from the projected
logical Hamiltonian derived in Eq.~(\ref{eq:logical_hamiltonian}), this Hamiltonian has the form of a single-qubit control Hamiltonian in the encoded subspace. The exchange coupling $J$ generates rotations about the logical $X_L$ axis, while the detuning $\Delta=\omega_1-\omega_2$ generates rotations about the logical $Z_L$ axis. These two control knobs provide the basis for constructing elementary encoded single-logical-qubit gates.

A pure logical $X_L$ rotation is generated when the exchange term is effectively absent during the phase-accumulation window. This condition can be realized in the projected model by setting an effective exchange rate $J_{\rm eff}=0$, or operationally by using a calibrated echo or logical-frame compensation that removes the unwanted $X_L$ component. Under this condition,
\begin{equation}
	H_X=\hbar JX_L
	\label{eq:HX}
\end{equation}
The resulting encoded rotation is
\begin{equation}
	R_X^L(\theta)=e^{-i\frac{\theta}{2}X_L},
	\qquad \theta=2JT
	\label{eq:RX_gate}
\end{equation}
A logical bit-flip operation corresponds to $\theta=\pi$, giving the gate
duration
\begin{equation}
	T_X^{(\pi)}=\frac{\pi}{2J}
	\label{eq:X_gate_time}
\end{equation}
In the physical two-qubit basis, this operation corresponds to the coherent single-excitation exchange $|01\rangle\leftrightarrow |10\rangle$.

In the detuning-dominated regime, the exchange contribution is either
suppressed, perturbatively small compared with $|\Delta|$, or compensated
in an effective rotating frame. The logical Hamiltonian is then approximated
by
\begin{equation}
	H_Z=\hbar\frac{\Delta}{2}Z_L
	\label{eq:HZ}
\end{equation}
The corresponding encoded phase rotation is
\begin{equation}
	R_Z^L(\theta)=e^{-i\frac{\theta}{2}Z_L},
	\qquad \theta=\Delta T
	\label{eq:RZ_gate}
\end{equation}
A logical $Z_L$ phase flip corresponds to $\theta=\pi$, with gate duration
\begin{equation}
	T_Z^{(\pi)}=\frac{\pi}{|\Delta|}
	\label{eq:TZpi}
\end{equation}
In the numerical benchmarks, an ideal $Z_L$ target gate should be interpreted as an exchange-compensated logical phase operation. If without compensation, the same control parameters produce a tilted-axis gate rather than a pure $Z_L$ gate. This convention is important for the parameter set in Table~\ref{tab:parameters}, where $J/2\pi=2.5MHz$ and $|\Delta|/2\pi=5MHz$. Since $J$ and $|\Delta|/2$ are comparable, the uncompensated rotation axis is not detuning dominated. When both $J$ and $\Delta$ are non-zero, the logical qubit rotates about a tilted axis,
\begin{equation}
	R_{\hat{n}}^L(\theta)
	=e^{-i\frac{\theta}{2}\hat{n}\cdot\vec{\sigma}_L}
	\label{eq:general_rotation}
\end{equation}
where
\begin{equation}
	\hat{n}
	=
	\frac{(J,0,\Delta/2)}
	{\sqrt{J^2+(\Delta/2)^2}},
	\qquad
	\theta
	=
	2T\sqrt{J^2+(\Delta/2)^2}
	\label{eq:rotation_axis_angle}
\end{equation}
Eqs.\eqref{eq:RX_gate}-\eqref{eq:rotation_axis_angle} show that arbitrary single-logical-qubit rotations can be constructed by controlling the exchange coupling and detuning.

Fig.\ref{fig:2}(b) summarizes the benchmarking protocol. We use two complementary quantities to evaluate the encoded logical gates: the average gate fidelity and the leakage probability. The average gate fidelity measures how close the noisy encoded operation $\mathcal{E}_L$ to the ideal target gate $U_L^{\mathrm{ideal}}$. A two dimensional logical subspace can be written as \cite{ABAD2022}
\begin{equation}
	F_{\mathrm{avg}}=\frac{dF_e+1}{d+1},\qquad d=2,
	\label{eq:Favg_section}
\end{equation}
where $F_e$ is the entanglement fidelity between the noisy logical operation and the ideal gate. A value of $F_{\mathrm{avg}}=1$ corresponds to a perfect encoded gate.

The leakage probability quantifies the population lost from the logical subspace during the gate operation. With the logical projector
\begin{equation}
	P_L
	=
	|01\rangle\langle 01|
	+
	|10\rangle\langle 10|,
	\label{eq:PL_gate_section}
\end{equation}
we define
\begin{equation}
	P_{\mathrm{leak}}(T)
	=
	1-
	\mathrm{Tr}
	\left[
	P_L\rho(T)P_L
	\right].
	\label{eq:Pleak_section}
\end{equation}
a high-fidelity logical rotation is meaningful only if the system remains inside the encoded manifold\cite{Hyyppa2024}. In the ideal projected model, $P_{\mathrm{leak}}=0$. However, in a realistic two-fluxonium device, higher-level admixture, non-adiabatic tuning, imperfect spectral isolation, control imperfections, and noise-induced transitions may populate states outside $\{|01\rangle,|10\rangle\}$.

The two metrics are used to quantify how correlated non-Markovian TLS noise affects encoded logical gates and whether the
odd-parity encoding improves the robustness of fluxonium-based logical control under control imperfections relevant to superconducting-qubit operations and dynamical-decoupling-based error mitigation \cite{Tripathi2022,Ezzell2023}.

\section{Numerical Results}

We numerically evaluate the performance of the odd-parity encoded logical gates under correlated longitudinal noise. In the following, we test whether the correlation-assisted suppression of differential noise derived in Sec.~II leads to a measurable gate-level advantage. In the previous sections, the odd-parity subspace can convert the two physical longitudinal fluctuations into a common-mode component and a differential component. The common-mode part contributes a global phase in the encoded subspace, the differential fluctuation $\delta\xi(t)=\xi_1(t)-\xi_2(t)$ produces a logical dephasing. The numerical analysis focuses on how the average gate fidelity changes as the spatial noise correlation.

In the presence of correlated TLS noise, the implemented operation is no longer generated solely by the ideal logical Hamiltonian. Within the projected logical model, the noisy evolution is governed by
\begin{equation}
	H_{\rm tot}^{(L)}(t)
	=
	H_L+H_{\rm noise}^{(L)}(t),
	\label{eq:HtotL}
\end{equation}
where $H_L$ generates the intended encoded rotation and
$H_{\rm noise}^{(L)}(t)$ contains the projected longitudinal noise. The
corresponding noisy logical operation is obtained by averaging over an
ensemble of stochastic trajectories, and the resulting logical map after a gate duration $T$ is written as
\begin{equation}
	\rho_L(T)=\mathcal{E}_L(T)\rho_L(0).
	\label{eq:noisy_map}
\end{equation}
Here, $\mathcal{E}_L$ captures the effect of differential TLS fluctuations on the encoded operation. The corresponding ideal target is
\begin{equation}
	\rho_L^{\rm ideal}(T)
	=
	U_L^{\rm ideal}(T)\rho_L(0)U_L^{{\rm ideal}\dagger}(T),
	\label{eq:ideal_map}
\end{equation}
where $U_L^{\rm ideal}$ can be chosen as $R_X^L$, $R_Z^L$, or a general rotation $R_{\hat n}^L$. In a full fluxonium implementation, the same comparison is supplemented by monitoring population outside the logical subspace.

This noisy operation is then compared with the ideal target gate $U_L^{\rm ideal}$ by using the average gate fidelity $F_{\rm avg}$. We use symmetric noise amplitudes $\sigma_1=\sigma_2=\sigma$, the differential-noise correlation function satisfy Eq.(21), which makes explicit that increasing $\rho$ suppresses the logical dephasing noise experienced by the encoded qubit.

\begin{figure}[htbp]
	\centering
	\includegraphics[width=8.6cm]{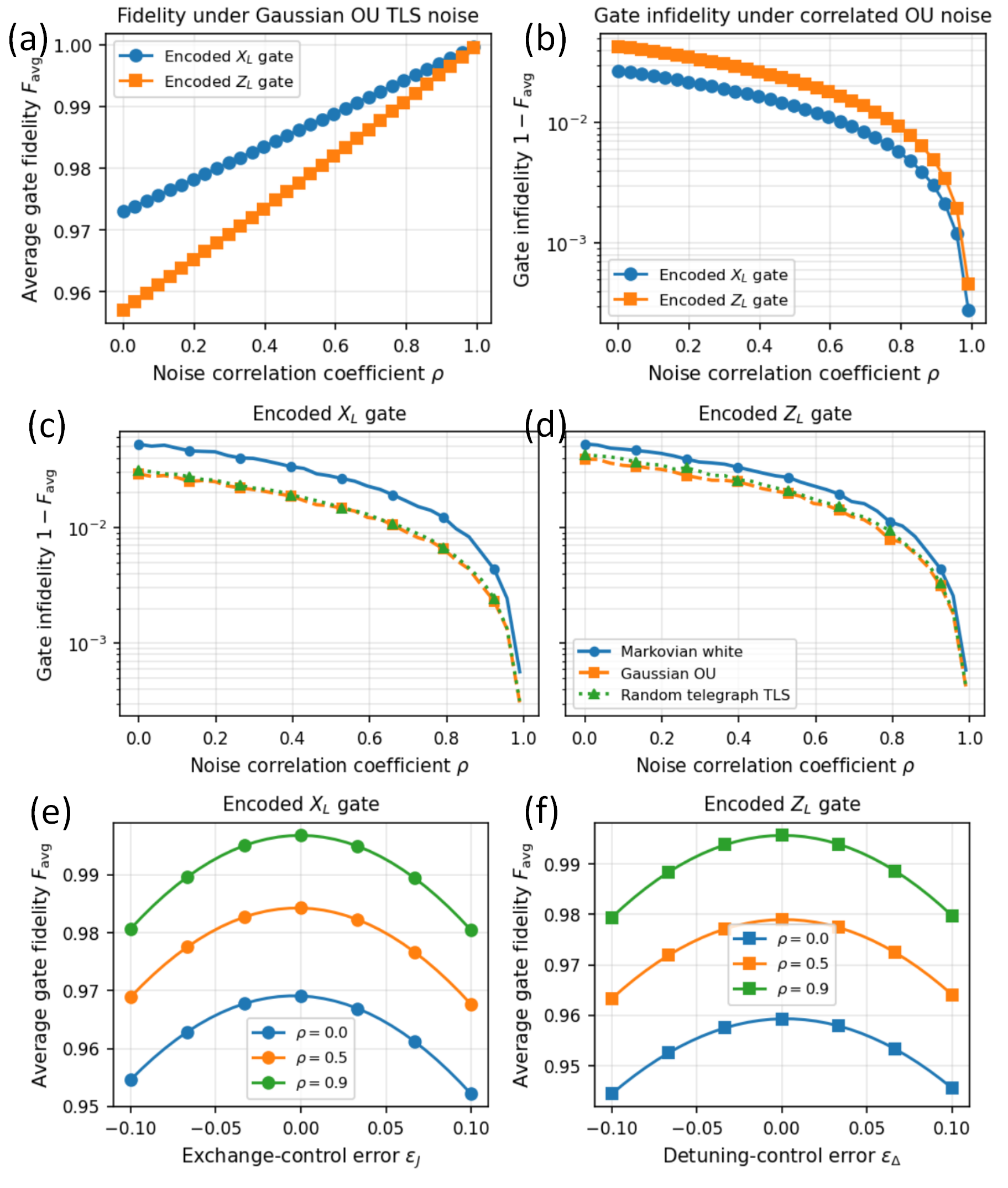}
	\caption{
		Numerical performance of odd-parity encoded logical gates under
		correlated longitudinal noise.
		(a) Average gate fidelity $F_{\rm avg}$ of the encoded $X_L$ and
		$Z_L$ gates as a function of the spatial noise-correlation coefficient
		$\rho$ under Gaussian Ornstein-Uhlenbeck (OU) TLS noise.
		(b) Corresponding gate infidelity $1-F_{\rm avg}$ on a logarithmic
		scale.
		(c) Gate infidelity of the encoded $X_L$ gate under Markovian white
		noise, Gaussian OU noise, and random-telegraph TLS noise.
		(d) Gate infidelity of the encoded $Z_L$ gate under the same three
		noise models.
		(e) Average gate fidelity of the encoded $X_L$ gate versus the
		exchange-control error $\epsilon_J$ for representative values of
		$\rho$.
		(f) Average gate fidelity of the encoded $Z_L$ gate versus the
		detuning-control error $\epsilon_\Delta$ for representative values of
		$\rho$.
	}
	\label{fig:3}
\end{figure}

The encoded $X_L$ gate is implemented in the exchange-dominated regime, with gate duration $T_X^{(\pi)}$ in Eq.(25). The encoded $Z_L$ gate is implemented in the detuning-dominated regime, with $T_Z^{(\pi)}$ in Eq.(28). The numerical simulations are performed in dimensionless units set by the logical control scale. The main results are summarized in Fig.~\ref{fig:3}.

We first consider the Gaussian OU TLS-noise model used in the theoretical
formulation. Fig.~\ref{fig:3}(a) shows the average gate
fidelity of the encoded $X_L$ and $Z_L$ gates as a function of the noise
correlation coefficient $\rho$. For both logical gates, the fidelity
increases monotonically as the noise becomes more spatially correlated.
This behavior directly follows from Eq.~\eqref{eq:numerical_differential_noise}.
When the two physical qubits experience largely independent fluctuations,
the differential noise remains strong and the encoded logical qubit
undergoes appreciable dephasing during the gate operation. As $\rho$
increases, the two fluctuations acquire a larger common component. This
common-mode component does not affect the relative phase between
$|01\rangle$ and $|10\rangle$, and the harmful differential fluctuation is
therefore reduced. As a result, the noisy operation $\mathcal{E}_L$ becomes
closer to the ideal target gate.

The same conclusion is emphasized in
Fig.~\ref{fig:3}(b), where the gate infidelity
$1-F_{\rm avg}$ is plotted on a logarithmic scale. This representation
makes the error suppression more visible. In the weakly correlated regime,
the encoded gate error is dominated by differential dephasing. In the
strongly correlated regime, the infidelity decreases rapidly because the
dominant longitudinal noise becomes nearly common-mode. Thus, the
correlation-assisted protection mechanism is not merely a state-preservation
effect. It directly improves the quality of encoded logical gates.

The encoded $X_L$ and $Z_L$ gates show the same qualitative trend, but their
absolute fidelities are not identical. This difference originates from the
different control Hamiltonians and gate durations associated with the two
operations. The $X_L$ gate is driven by the exchange term, while the $Z_L$
gate is generated by detuning. Therefore, the accumulated stochastic phase
and the sensitivity to longitudinal fluctuations can differ quantitatively.
Nevertheless, both gates benefit from increasing $\rho$, confirming that
the advantage is a property of the odd-parity encoding rather than a
special feature of a single gate type.

We next examine whether the correlation-assisted advantage is specific to
the Gaussian OU noise model. To this end, we compare three representative
longitudinal dephasing-noise models: Markovian white noise, Gaussian OU
colored noise, and random-telegraph TLS noise. The white-noise model
represents a memoryless dephasing process with short temporal correlation.
The OU model represents Gaussian colored noise with a finite memory time
$\tau_c$. The random-telegraph model represents a non-Gaussian TLS-like
fluctuator with discrete switching dynamics. These models differ in their
temporal statistics and spectra, but all generate longitudinal fluctuations
that enter the encoded logical Hamiltonian through the differential-noise
channel. For a fair comparison, the parameters of the three models are
chosen so that their effective single-qubit dephasing strengths are
comparable.

Figs.~\ref{fig:3}(c) and \ref{fig:3}(d) show the gate infidelity for the encoded $X_L$ and $Z_L$ gates under these three noise models. Although the absolute value of the infidelity depends on the noise spectrum and temporal
statistics, all models exhibit the same overall trend:
\begin{equation}
	\rho \uparrow
	\quad \Rightarrow \quad
	1-F_{\rm avg} \downarrow .
\end{equation}
This result is important because it shows that the fidelity improvement is
not an artifact of the OU assumption. Instead, it follows from the
projection of correlated longitudinal noise into the odd-parity logical
subspace. Once the noise becomes more spatially correlated, the
differential component decreases, regardless of whether the microscopic
fluctuation is memoryless, Gaussian colored, or random telegraph-like.
Therefore, the encoded logical gate inherits a noise-filtering property
from the symmetry of the subspace.

Finally, we study the robustness of the encoded gates against coherent
control imperfections. In an experiment, the exchange coupling and detuning
cannot be calibrated perfectly. We model these imperfections as
\begin{equation}
	J \rightarrow J(1+\epsilon_J),
	\qquad
	\Delta \rightarrow \Delta(1+\epsilon_\Delta),
	\label{eq:control_errors}
\end{equation}
where $\epsilon_J$ and $\epsilon_\Delta$ denote the fractional control
errors. The exchange error mainly affects the encoded $X_L$ gate by
changing the rotation rate and therefore the final rotation angle. The
detuning error similarly affects the encoded $Z_L$ gate by modifying the
accumulated logical phase.

Fig.~\ref{fig:3}(e) shows the average fidelity of the
encoded $X_L$ gate as a function of $\epsilon_J$ for several values of
$\rho$. The fidelity is maximized near $\epsilon_J=0$ and decreases as the
magnitude of the exchange-control error increases. This behavior reflects
the coherent over-rotation or under-rotation induced by imperfect exchange
calibration. However, at any fixed value of $\epsilon_J$, the gate with
larger noise correlation has higher fidelity. In particular, the
$\rho=0.9$ curve remains well above the $\rho=0.5$ and $\rho=0$ curves
throughout the error range considered.

Fig.~\ref{fig:3}(f) shows the corresponding result for
the encoded $Z_L$ gate under detuning-control error. The same qualitative
behavior is observed: detuning miscalibration reduces the fidelity, but
larger spatial noise correlation still improves the gate performance. These
results show that correlation-assisted protection and control calibration
address complementary error mechanisms. Spatial correlation suppresses
stochastic logical dephasing, while accurate calibration suppresses
coherent rotation errors. Therefore, correlated noise does not remove the
need for precise control, but it provides an additional robustness channel
for encoded logical operations.

The numerical simulations in this section are restricted to the projected two-dimensional logical subspace. Since the reduced model contains only the encoded basis states, population transfer to states outside the logical manifold is not explicitly evaluated here. The results in Fig.~3 should therefore be interpreted as logical-channel fidelities within the encoded subspace, rather than complete device-level gate fidelities.

This distinction is important. In a full multilevel fluxonium simulation, higher-level admixture, nonadiabatic parameter tuning, imperfect spectral isolation, transverse noise components, and pulse-shape constraints can populate states outside the logical manifold. Those effects are not included in the present projected analysis. The role of the present section is instead to isolate the encoded-subspace mechanism: positive spatial correlation suppresses the differential longitudinal fluctuation and thereby improves the fidelity of odd-parity encoded logical gates.

Within this scope, the numerical results show three main points. First, increasing $\rho$ improves the average fidelity of both exchange-based $X_L$ gates and detuning-based $Z_L$ gates. Second, this improvement persists across Gaussian OU, Markovian white, and random-telegraph longitudinal-noise models. Third, the correlation-assisted advantage remains visible in the presence of exchange- and detuning-control errors, although accurate calibration is still required to suppress coherent rotation errors.

\section{Dynamical-Decoupling-Assisted Logical Gates}

The numerical results in Sec.~IV show that spatially correlated longitudinal TLS noise improves the fidelity of odd-parity encoded logical gates by suppressing the differential fluctuation between the two physical qubits. However, this protection is not complete when the spatial correlation is imperfect. For $\rho<1$, a residual differential fluctuation remains and acts as an effective logical dephasing channel in the encoded subspace. In this section, we examine whether logical dynamical decoupling (DD) can further reduce this residual dephasing during encoded gate operations.

The role of DD here should be understood at the logical level. Instead of modeling a full pulse-level implementation on the multilevel fluxonium Hilbert space, we represent the refocusing operations directly in the projected encoded subspace. Logical $\pi$ pulses are inserted so that the sign of the residual dephasing term is periodically reversed. For slowly varying TLS fluctuations, this sign reversal partially cancels the stochastic phase accumulated during the gate. Thus, DD acts as a second layer of protection: spatial noise correlation reduces the differential noise at its source, while logical refocusing suppresses the remaining low-frequency differential component.

\begin{figure}[t]
	\centering
	\includegraphics[width=6.8cm]{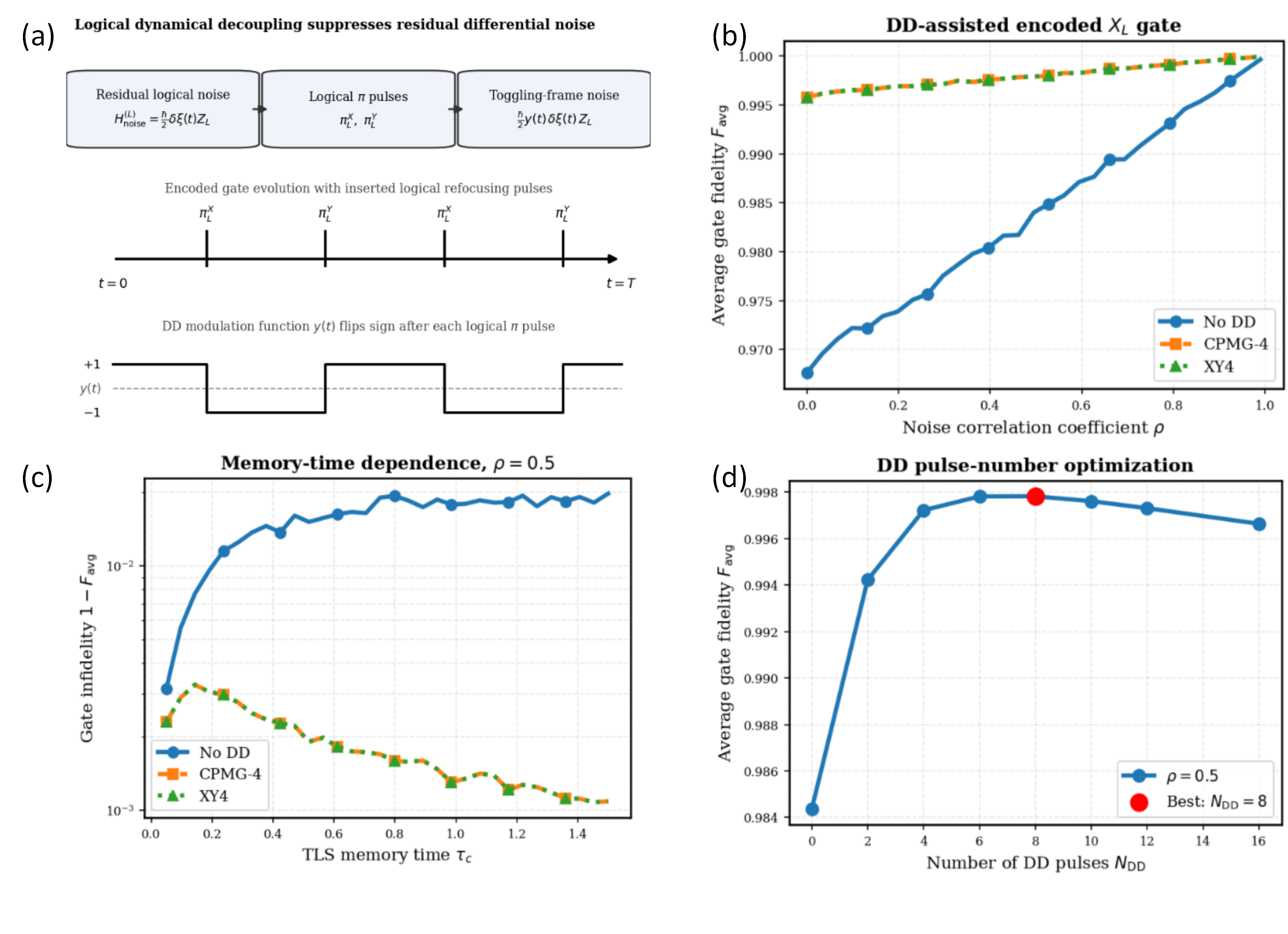}
\caption{
	Dynamical-decoupling-assisted encoded logical gates in the projected odd-parity subspace.
	(a) Schematic of logical refocusing, where inserted logical $\pi$ pulses reverse the sign of the residual differential dephasing during the encoded gate.
	(b) Average fidelity of the encoded $X_L$ gate as a function of the spatial noise-correlation coefficient $\rho$ with no DD, CPMG-4, and XY4.
	(c) Gate infidelity as a function of the TLS memory time at fixed intermediate correlation, showing that DD is most effective against slowly varying non-Markovian differential noise.
	(d) Average fidelity versus the number of DD pulses, illustrating the tradeoff between improved noise averaging and pulse-induced control errors.
}

	\label{fig4}
\end{figure}

Fig.~4(a) illustrates this mechanism schematically. The logical refocusing pulses are not treated as independent operations simply added after the gate. Rather, the target encoded rotation and the refocusing sequence are regarded as a compensated logical-control protocol. In this picture, the desired logical gate is preserved, while the residual differential dephasing is averaged over the gate duration. This idealized treatment is useful for isolating the noise-filtering effect of DD, but it does not include the full cost of implementing the refocusing pulses in a realistic multilevel fluxonium device.

We first compare the encoded $X_L$ gate fidelity with no DD, CPMG-4, and XY4 sequences. As shown in Fig.~4(b), all three cases benefit from increasing spatial noise correlation, consistent with the results in Sec.~IV. When DD is applied, the fidelity is further enhanced over a broad range of $\rho$. The improvement is most visible at weak and intermediate correlations, where the residual differential noise is still appreciable. In the strongly correlated regime, the differential noise has already been largely converted into a harmless common-mode fluctuation, so the additional gain from DD becomes smaller.

This behavior clarifies the relation between correlation-assisted encoding and DD. The two mechanisms do not address exactly the same level of the problem. Spatial correlation changes the effective noise seen by the encoded qubit by suppressing the differential component. DD then acts on whatever residual differential noise remains. Therefore, DD is most useful when the noise is partially correlated but not yet close to the common-mode limit.

The benefit of DD is especially clear for non-Markovian TLS noise with a finite memory time. Fig.~4(c) shows the gate infidelity as a function of the TLS memory time at fixed intermediate correlation. Without DD, slowly varying differential fluctuations can accumulate coherent phase errors during the gate, leading to larger infidelity. With CPMG-4 or XY4, these slow fluctuations are more effectively refocused, and the gate error is reduced. This indicates that logical DD is particularly helpful when the residual TLS noise has appreciable temporal correlation rather than being purely memoryless.

Finally, Fig.~4(d) examines the dependence on the number of refocusing pulses. Increasing the number of DD pulses initially improves the fidelity because the residual differential noise is averaged more efficiently. However, the improvement eventually saturates and can decrease when pulse imperfections are included. This reflects a practical tradeoff. Too few pulses do not sufficiently suppress slow differential noise, whereas too many pulses introduce additional opportunities for control errors. The DD pulse number should therefore be optimized rather than increased indefinitely.

Overall, these results show that logical DD can complement correlation-assisted odd-parity encoding. Spatial correlation suppresses the dominant differential longitudinal noise, and DD further filters the residual slow component during the encoded gate. Within the projected logical model, this combination improves the logical-channel fidelity of encoded gates under correlated non-Markovian TLS noise. A complete device-level assessment would require multilevel simulations including leakage, pulse-shape constraints, finite pulse duration, and calibration errors associated with the physical implementation of the logical refocusing operations.

\section{Conclusion}

In this work, we proposed and analyzed odd-parity encoded logical gates in
two coupled fluxonium qubits subject to correlated non-Markovian TLS noise.
By projecting the two-fluxonium Hamiltonian onto the subspace spanned by
$|01\rangle$ and $|10\rangle$, we obtained an effective logical Hamiltonian
in which the exchange coupling generates $X_L$ rotations and the qubit
detuning generates $Z_L$ rotations. The same projection shows that
common-mode longitudinal TLS noise contributes only a global phase, while
the harmful component is the differential fluctuation
$\delta\xi(t)=\xi_1(t)-\xi_2(t)$.

Numerical simulations show that increasing the spatial noise correlation
systematically improves the average gate fidelity of encoded logical
operations. This enhancement persists across several representative
dephasing-noise models, including Markovian white noise, Gaussian
Ornstein-Uhlenbeck noise, and random-telegraph TLS noise. We further
showed that the encoded gates remain more robust at high noise correlation
even in the presence of exchange- and detuning-control errors. Finally,
logical dynamical decoupling provides an additional layer of protection by
filtering residual differential $Z_L$ dephasing during the gate operation.

These findings suggest that correlated longitudinal noise, usually regarded as a source of decoherence, can be made less harmful by using a symmetry-adapted logical encoding.

\begin{acknowledgments}
		This work was funded by the State Key Laboratory of Quantum Optics Technologies and Devices, Shanxi University, Shanxi, China (Grants No.KF202503); Zhejiang Key Laboratory of Quantum State Control and Optical Field Manipulation, Hangzhou Dianzi
        University (KYZ074326001).
	\end{acknowledgments}
    
\section*{DATA AVAILABILITY}
The data that support the findings of this article are not publicly available. The data are available from the authors upon reasonable request.

\bibliography{ref}

@article{bao2022,
	title={Fluxonium: An alternative qubit platform for high-fidelity operations},
	author={Bao, Feng and Deng, Hao and Ding, Dawei and Gao, Ran and Gao, Xun and Huang, Cupjin and Jiang, Xun and Ku, Hsiang Sheng and Li, Zhisheng and Ma, Xizheng and others},
	journal={Physical Review Letters},
	volume={129},
	number={1},
	pages={010502},
	year={2022},
	publisher={APS}
}

@article{somoroff2023,
title={Millisecond coherence in a superconducting qubit},
author={Somoroff, Aaron and Ficheux, Quentin and Mencia, Raymond A and Xiong, Haonan and Kuzmin, Roman and Manucharyan, Vladimir E},
journal={Physical Review Letters},
volume={130},
number={26},
pages={267001},
year={2023},
publisher={APS}
}

@article{nguyen2022,
title={Blueprint for a high-performance fluxonium quantum processor},
author={Nguyen, Long B and Koolstra, Gerwin and Kim, Yosep and Morvan, Alexis and Chistolini, Trevor and Singh, Shraddha and Nesterov, Konstantin N and J{\"u}nger, Christian and Chen, Larry and Pedramrazi, Zahra and others},
journal={PRX Quantum},
volume={3},
number={3},
pages={037001},
year={2022},
publisher={APS}
}

@article{moskalenko2022,
title={High fidelity two-qubit gates on fluxoniums using a tunable coupler},
author={Moskalenko, Ilya N and Simakov, Ilya A and Abramov, Nikolay N and Grigorev, Alexander A and Moskalev, Dmitry O and Pishchimova, Anastasiya A and Smirnov, Nikita S and Zikiy, Evgeniy V and Rodionov, Ilya A and Besedin, Ilya S},
journal={Npj Quantum Information},
volume={8},
number={1},
pages={130},
year={2022},
publisher={Nature Publishing Group UK London}
}

@article{zhang2024,
title={Tunable inductive coupler for high-fidelity gates between fluxonium qubits},
author={Zhang, Helin and Ding, Chunyang and Weiss, DK and Huang, Ziwen and Ma, Yuwei and Guinn, Charles and Sussman, Sara and Chitta, Sai Pavan and Chen, Danyang and Houck, Andrew A and others},
journal={PRX Quantum},
volume={5},
number={2},
pages={020326},
year={2024},
publisher={APS}
}

@article{rosenfeld2024,
title={High-fidelity two-qubit gates between fluxonium qubits with a resonator coupler},
author={Rosenfeld, Emma L and Hann, Connor T and Schuster, David I and Matheny, Matthew H and Clerk, Aashish A},
journal={PRX Quantum},
volume={5},
number={4},
pages={040317},
year={2024},
publisher={APS}
}

@article{mencia2024,
title={Integer fluxonium qubit},
author={Mencia, Raymond A and Lin, Wei Ju and Cho, Hyunheung and Vavilov, Maxim G and Manucharyan, Vladimir E},
journal={PRX Quantum},
volume={5},
number={4},
pages={040318},
year={2024},
publisher={APS}
}

@article{lin2025,
title={24 days-stable CNOT gate on fluxonium qubits with over 99.9\% fidelity},
author={Lin, Wei Ju and Cho, Hyunheung and Chen, Yinqi and Vavilov, Maxim G and Wang, Chen and Manucharyan, Vladimir E},
journal={PRX Quantum},
volume={6},
number={1},
pages={010349},
year={2025},
publisher={APS}
}

@article{deGraaf2021,
title={Quantifying dynamics and interactions of individual spurious low-energy fluctuators in superconducting circuits},
author={De Graaf, SE and Mahashabde, Sumedh and Kubatkin, SE and Tzalenchuk, A Ya and Danilov, AV},
journal={Physical Review B},
volume={103},
number={17},
pages={174103},
year={2021},
publisher={APS}
}

@article{abdurakhimov2022,
title={Identification of different types of high-frequency defects in superconducting qubits},
author={Abdurakhimov, Leonid V and Mahboob, Imran and Toida, Hiraku and Kakuyanagi, Kosuke and Matsuzaki, Yuichiro and Saito, Shiro},
journal={PRX Quantum},
volume={3},
number={4},
pages={040332},
year={2022},
publisher={APS}
}

@article{chen2024,
title={Phonon engineering of atomic-scale defects in superconducting quantum circuits},
author={Chen, Mo and Owens, John Clai and Putterman, Harald and Sch{\"a}fer, Max and Painter, Oskar},
journal={Science Advances},
volume={10},
number={37},
pages={eado6240},
year={2024},
publisher={American Association for the Advancement of Science}
}

@article{odeh2025,
title={Non-Markovian dynamics of a superconducting qubit in a phononic bandgap},
author={Odeh, Mutasem and Godeneli, Kadircan and Li, Eric and Tangirala, Rohin and Zhou, Haoxin and Zhang, Xueyue and Zhang, Zi Huai and Sipahigil, Alp},
journal={Nature Physics},
volume={21},
number={3},
pages={406-411},
year={2025},
publisher={Nature Publishing Group UK London}
}

@article{zhuang2026,
title={Non-Markovian relaxation rpectroscopy of fluxonium qubits},
author={Zhuang, Ze Tong and Rosenstock, Dario and Liu, Bao Jie and Somoroff, Aaron and Manucharyan, Vladimir E and Wang, Chen},
journal={Nature Communications},
volume={17},
pages={3209},
year={2026},
publisher={Nature Publishing Group UK London}
}

@article{zhang2022,
title={Predicting non-markovian superconducting-qubit dynamics from tomographic reconstruction},
author={Zhang, Haimeng and Pokharel, Bibek and Levenson-Falk, EM and Lidar, Daniel},
journal={Physical Review Applied},
volume={17},
number={5},
pages={054018},
year={2022},
publisher={APS}
}

@article{white2022,
title={Non-Markovian quantum process tomography},
author={White, Gregory AL and Pollock, Felix A and Hollenberg, Lloyd CL and Modi, Kavan and Hill, Charles D},
journal={PRX Quantum},
volume={3},
number={2},
pages={020344},
year={2022},
publisher={APS}
}

@article{gulacsi2023,
title={Signatures of non-Markovianity of a superconducting qubit},
author={Gul{\'a}csi, Bal{\'a}zs and Burkard, Guido},
journal={Physical Review B},
volume={107},
number={17},
pages={174511},
year={2023},
publisher={APS}
}

@article{zou2024,
title={Spatially correlated classical and quantum noise in driven qubits},
author={Zou, Ji and Bosco, Stefano and Loss, Daniel},
journal={Npj Quantum Information},
volume={10},
number={1},
pages={46},
year={2024},
publisher={Nature Publishing Group UK London}
}

@article{zanardi1997,
title={Noiseless quantum codes},
author={Zanardi, Paolo and Rasetti, Mario},
journal={Physical Review Letters},
volume={79},
number={17},
pages={3306},
year={1997},
publisher={APS}
}

@article{lidar1998,
title={Decoherence free subspaces for quantum computation},
author={Lidar, Daniel A and Chuang, Isaac L and Whaley, K Birgitta},
 journal = {Physics Review Letters},
    volume = {81},
    pages = {2594},
   year={1998}
}

@article{ji2026,
    author = {Ji, Chenghong and Zhao, Chaoying},
    title = {Entanglement dynamics of multi-fluxonium-qubits under non-Markovian TLS noise},
    journal = {Applied Physics Letters},
    volume = {128},
    number = {20},
    pages = {202601},
    year = {2026},
}

@article{ABAD2022,
  title = {Universal Fidelity Reduction of Quantum Operations from Weak Dissipation},
  author = {Abad, Tahereh and Fern\'andez-Pend\'as, Jorge and Frisk Kockum, Anton and Johansson, G\"oran},
  journal = {Physics Review Letters},
  volume = {129},
  issue = {15},
  pages = {150504},
  year = {2022},
  publisher = {American Physical Society},
}

@article{manucharyan2009,
title={Fluxonium: Single cooper-pair circuit free of charge offsets},
author={Manucharyan, Vladimir E and Koch, Jens and Glazman, Leonid I and Devoret, Michel H},
journal={Science},
volume={326},
number={5949},
pages={113-116},
year={2009},
publisher={American Association for the Advancement of Science}
}

@article{manucharyan2009fluxonium,
title={Fluxonium: Single cooper-pair circuit free of charge offsets},
author={Manucharyan, Vladimir E and Koch, Jens and Glazman, Leonid I and Devoret, Michel H},
journal={Science},
volume={326},
number={5949},
pages={113-116},
year={2009},
publisher={American Association for the Advancement of Science}
}

@article{blais2021circuit,
title={Circuit quantum electrodynamics},
author={Blais, Alexandre and Grimsmo, Arne L and Girvin, Steven M and Wallraff, Andreas},
journal={Reviews of Modern Physics},
volume={93},
number={2},
pages={025005},
year={2021},
publisher={APS}
}

@article{ithier2005decoherence,
title={Decoherence in a superconducting quantum bit circuit},
author={Ithier, Gregoire and Collin, E and Joyez, P and Meeson, PJ and Vion, Denis and Esteve, Daniel and Chiarello, F and Shnirman, A and Makhlin, Yu and Schriefl, Josef and others},
journal={Physical Review B},
volume={72},
number={13},
pages={134519},
year={2005},
publisher={APS}
}

@article{wang1945theory,
title={On the theory of the Brownian motion II},
author={Wang, Ming Chen and Uhlenbeck, George Eugene},
journal={Reviews of Modern Physics},
volume={17},
number={2-3},
pages={323},
year={1945},
publisher={APS}
}

@article{hyyppa2024,
title={Reducing leakage of single-qubit gates for superconducting quantum processors using analytical control pulse envelopes},
author={Hyypp{\"a}, Eric and Veps{\"a}l{\"a}inen, Antti and Papi{\v{c}}, Miha and Chan, Chun Fai and Inel, Sinan and Landra, Alessandro and Liu, Wei and Luus, J{\"u}rgen and Marxer, Fabian and Ockeloen Korppi, Caspar and others},
journal={PRX Quantum},
volume={5},
number={3},
pages={030353},
year={2024},
publisher={APS}
}

@article{tripathi2022,
title={Suppression of crosstalk in superconducting qubits using dynamical decoupling},
author={Tripathi, Vinay and Chen, Huo and Khezri, Mostafa and Yip, Ka-Wa and Levenson Falk, EM and Lidar, Daniel A},
journal={Physical Review Applied},
volume={18},
number={2},
pages={024068},
year={2022},
publisher={APS}
}

@article{ezzell2023,
title={Dynamical decoupling for superconducting qubits: A performance survey},
author={Ezzell, Nic and Pokharel, Bibek and Tewala, Lina and Quiroz, Gregory and Lidar, Daniel A},
journal={Physical Review Applied},
volume={20},
number={6},
pages={064027},
year={2023},
publisher={APS}
}

\end{document}